\documentclass[12pt]{article}
\pdfoutput=1

\usepackage[english]{babel}
\usepackage{amssymb,amsmath}
\usepackage{amsfonts}
\usepackage{latexsym}
\usepackage{verbatim}
\usepackage{amsthm}
\usepackage{amsbsy}
\usepackage{graphicx,color}
\usepackage{epsfig}
\usepackage{verbatim}
\usepackage{cite}

\newcommand{\be}{\begin{equation}}
\newcommand{\ee}{\end{equation}}
\newcommand{\bi}{\begin{itemize}}
\newcommand{\ei}{\end{itemize}}
\newcommand{\bea}{\begin{eqnarray}}
\newcommand{\eea}{\end{eqnarray}}
\newcommand{\ba}{\begin{array}}
\newcommand{\ea}{\end{array}}

\def\abar{{\bar a}}

\def\mubar{{\bar \mu}}

\def\sigmabar{{\bar\sigma}}
\def\ibar{{\bar\imath}}
\def\jbar{{\bar\jmath}}
\def\kbar{{\bar k}}

\def\mbar{{\bar m}}

\def\partialbar{{\bar\partial}}

\newcommand{\newsection}[1]{\section{#1}\setcounter{equation}{0}}
\newcommand{\nn}{\nonumber}
\newlength{\sswidth}

\usepackage[debug,pageanchor=false]{hyperref}
\definecolor{link}{rgb}{.8,.15,.1}
\hypersetup{colorlinks=true,linkcolor=link,citecolor=link,urlcolor=link,linktocpage}

\begin{document}
\begin{titlepage}

\vskip 2cm
\begin{center}
{\Large \textbf{B-strings on non-K\"ahlerian manifolds}} 
\vskip0.45cm

\end{center}
\vspace{1cm}

\vspace{1cm}

\begin{center}
  \textsc{ ~Camillo Imbimbo$^{1,2,a}$ 
  }  
\end{center}

\vspace{0.5cm}

\begin{center}
\sl
$^1\,$ Dipartimento di Fisica, Universit\`a di Genova,
Via Dodecaneso 33, 16146, Genoa, \rm ITALY \\
\sl $^2\,$ INFN, Sezione di Genova, Via Dodecaneso 33, 16146, Genoa, \rm ITALY

\vspace{0.3 cm}

 \vspace{0.3 cm}
\sl
 
\vspace{0.3 cm}
\end{center}

\begin{center}
{\small $^a$camillo.imbimbo@ge.infn.it 
}
\end{center}

\vspace{1cm}

\centerline{\textsc{ Abstract}}
 \vspace{0.2cm}
 
{\small  
We explain how to couple topological B-models whose targets are non-K\"ahlerian manifolds to topological
gravity and to thus define corresponding topological strings. We emphasize the need to take into account
the coupling to the superghost field of topological gravity in order to obtain a consistent definition of the
string model. We also review the importance of the superghost for correctly interpreting  the
holomorphic anomaly of the string amplitudes. We perform our analysis in the BV framework in
order to make it completely gauge independent.
 }
\vspace{0.2cm}
\thispagestyle{empty}

\vfill
\eject
\end{titlepage}
\hypersetup{pageanchor=true}
\baselineskip18pt
\tableofcontents
\newpage
\newsection{Introduction} 
\label{sec:intro}

The original formulation of topological gravity \cite{Labastida:1988zb} is deceptively simple:  it is characterized by a BRST operator
which acts  as an exterior differential on the space of space-time metrics
\bea
S_0 \, g_{\alpha\beta} = \psi_{\alpha\beta} \qquad S_0 \,  \psi_{\alpha\beta} = 0
\label{BRSTtrivial}
\eea
where $g_{\alpha\beta}$ is a space-time Riemannian metric and $\psi_{\alpha\beta}$ the topological gravitino field.

However, if Eq. (\ref{BRSTtrivial})  were all, topological gravity would have no  physical content, since the {\it local} BRST local cohomology of such nilpotent
transformations is obviously empty.  Raymond Stora \cite{Ouvry:1988mm} (together with others\cite{Baulieu:1988xs}),  had the crucial insight which clarified the physical meaning of the theory. He understood that the relevant notion  for topological gravity was local BRST cohomology  {\it equivariant} with respect to diffeomorphisms. The definition of the equivariant BRST operator requires introducing  the reparametrization ghost fields $c^\alpha$  of  ghost number +1 together with the ghost-for-ghost, or {\it superghost} field $\gamma^\alpha$  of ghost number +2. The equivariant nilpotent BRST transformations which, according to Raymond,  must replace (\ref{BRSTtrivial}) are then
\bea
& s\, g_{\alpha\beta} = -{\cal L}_c \,g_{\alpha\beta}+ \psi_{\alpha\beta} \qquad  &s\, \psi_{\alpha\beta} = -{\cal L}_c \,\psi_{\alpha\beta}+ {\cal L}_\gamma\,g_{\alpha\beta} \nn\\
& s\, c^\alpha = -\frac{1}{2}\,{\cal L}_c \, c^\alpha +\gamma^\alpha\qquad\quad&s\, \gamma^\alpha =- {\cal L}_c\,\gamma^\alpha 
\label{sequivariant}
\eea
 ${\cal L}_c$ is the Lie derivative implementing reparametrizations
associated to the vector field $c^\alpha$.  

The usual local cohomology of the BRST operator $s$ is just as trivial as that of the simple minded $S_0$ (\ref{BRSTtrivial}).  However
the  {\it equivariant} cohomology of $s$,  which is  the
local cohomology of $s$ on the space of field functionals which do {\it not} include the reparametrization ghost $c^\alpha$,  is not: this is the cohomology which  characterizes  the physical observables of the theory.
 
When working with the equivariant cohomology   it is convenient to define the operator 
\bea
S= s +{\cal L}_c
\eea
which acts as follows
\bea
&& S\, g_{\alpha\beta} = \psi_{\alpha\beta} \qquad S\, \psi_{\alpha\beta} = {\cal L}_\gamma\,g_{\alpha\beta} \qquad S\, \gamma^\alpha =0
\eea
Nilpotency of $s$ is equivalent to 
\bea
S^2 = {\cal L}_\gamma
\label{nilone}
\eea
on the space of fields  $g_{\alpha\beta}, \psi_{\alpha\beta}$ and $\gamma^\alpha$. The equivariant cohomology of $s$ is therefore the same  as the cohomology of $S$ on {\it reparametrization invariant functionals} of the fields $g_{\alpha\beta}, \psi_{\alpha\beta}$ and $\gamma^\alpha$ . It is useful to decompose $S$  as the sum of two nilpotent operators
\bea
S = S_0 + G_{\gamma}
\eea
where
\bea
&& S_0\, g_{\alpha\beta} = \psi_{\alpha\beta}\qquad S_0 \psi_{\alpha\beta}= 0\qquad S_0 \,\gamma^\alpha=0 \nn\\
&& G_\gamma \, g_{\alpha\beta} =0 \qquad G_\gamma\, \psi_{\alpha\beta}= {\cal L}_\gamma\,g_{\alpha\beta} \qquad G_\gamma \, \gamma^\alpha=0
\label{Sequivariant}
\eea
Eq. (\ref{nilone}) is equivalent to the super-algebra
\bea
S_0^2=G_\gamma^2 =0 \qquad \{ S_0, G_\gamma\}={\cal L}_\gamma
\label{algebrabrst}
\eea
$S_0$ is the ``naive'' topological gravity BRST operator (\ref{BRSTtrivial}) whose  local cohomology is empty. The $G_{\gamma}$ part of the BRST transformations,  linear in the superghost $\gamma^\alpha$,  provides the extension of $S_0$ to the equivariant, non-trivial,  $S$. 

In two space-time dimensions one can couple topological gravity to topological matter and define in this way {\it topological strings}  \cite{Montano:1988dr}---  much in the same way as ordinary (super)strings are defined by coupling 2-dimensional (super)gravity to (super)conformal matter field theories. 

Topological matter theories are characterized by nilpotent  BRST operators $S^{matter}_0$.  Coupling the topological matter field theory to topological gravity means to extend $S^{matter}_0$ to the gravity sector, in accordance to (\ref{sequivariant}),  by including  diffeomorphisms acting  on the matter fields 
\bea
s = -{\cal L}_c + S_0 +G_\gamma
\label{equivariantextension}
\eea
Nilpotency of $s$ on the matter fields requires adding  to $S^{matter}_0$ a piece, denoted by   $G^{matter}_\gamma$,  which is linear in the superghost $\gamma^\alpha$ and which satisfies, together with  $S^{matter}_0$,  the same super-algebra (\ref{algebrabrst}) which holds in the gravity sector.

From what we just said, it is apparent that the superghost dependent part of the equivariant BRST transformations is, from the algebraic point of view,  the crucial ingredient necessary for the consistent, equivariant,  coupling of topological matter to topological gravity. It is hence curious that, as a matter of fact, the superghost $\gamma^\alpha$ rarely makes its appearance, explicitly,   in the immense literature devoted to topological strings. To understand why, we need to recall the general features of the topological strings construction.

The action of topological matter quantum field theory has the form
\bea
\Gamma^{matter} [\Phi, g_{\alpha\beta}] = \Gamma_0[\Phi]+  \int S^{matter}_0 \Psi[\Phi, g_{\alpha\beta}]
\label{mattertftone}
\eea
where $\Phi$ denotes schematically the collection of matter fields; $\Gamma_0[\Phi] $ is both $S^{matter}_0$ invariant and invariant under space-time diffeomorphisms without the help of a space-time metric ---  i.e. it is a topological term.  In certain cases one can  take the ``classical'' term $\Gamma_0[\Phi] $ to vanish: this happens
when the local cohomology of the matter BRST operator is empty.  The action for these theories --- which are called
of  ``cohomological'' type ---  reduces to a pure gauge-fixing term. Semi-classical approximation
is exact for cohomological theories and we will restrict, for simplicity,  the following discussion to this class of topological
theories.

The second term in the action (\ref{mattertftone}) is
a gauge-fixing term: the gauge fermion $\Psi[\Phi, g_{\alpha\beta}]$ is arbitrary as long as it provides non-degenerate
kinetic terms for all the matter fields. To this end, it necessarily depends on a background space-time metric $ g_{\alpha\beta}$. For matter topological quantum field theories  $ g_{\alpha\beta}$ plays the role of a gauge-fixing parameter: the physics, thanks to the nilpotency of $ S^{matter}_0$, does not depend on the specific choice for $ g_{\alpha\beta}$.

To construct the topological string model based on a given matter topological QFT one needs to extend
$S^{matter}_0$ to the gravity sector in the equivariant way, as prescribed in (\ref{equivariantextension}). For cohomological theories, the action of the coupled system takes therefore the form
\bea
&&\Gamma^{mat+t.g.} [\Phi, g_{\alpha\beta}, \psi_{\alpha\beta}, \gamma^\alpha] =  \int S\, \Psi[\Phi, g_{\alpha\beta}]= \nn\\
&&\qquad\qquad =\Gamma^{matter} [\Phi, g_{\alpha\beta}] + \int \psi^{\alpha\beta} \, S_{\alpha\beta}[ \Phi, g_{\alpha\beta}] + \int G_\gamma\, \Psi[\Phi, g_{\alpha\beta}]
\label{mattgaction}
\eea
where $S_{\alpha\beta}$ is the topological super-current
\bea
S_{\alpha\beta} = \frac{\delta \, \Psi[\Phi, g_{\alpha\beta}]}{\delta g^{\alpha\beta}}
\label{supercurrentone}
\eea
From (\ref{mattgaction}) we see that topological matter couples to the topological gravity multiplet not only via  the super-current and the gravitino field $\psi^{\alpha\beta}$, but also  by means of terms proportional to the superghost $\gamma^\alpha$. 

The partition function obtained by integrating the matter fields
\bea
Z[g_{\alpha\beta}, \psi_{\alpha\beta}, \gamma^\alpha] = \int [d\,\Phi] \, \mathrm{e}^{-\Gamma^{mat+t.g.} [\Phi, g_{\alpha\beta}, \psi_{\alpha\beta}, \gamma^\alpha]}
\eea
is a functional of the topological gravity multiplet which satisfies the BRST identity
\bea
S\, Z[g_{\alpha\beta}, \psi_{\alpha\beta}, \gamma^\alpha] =0
\label{BRSTPartition}
\eea
This identity says that $Z[g_{\alpha\beta}, \psi_{\alpha\beta}, \gamma^\alpha] $ is an equivariant closed form
on the space of space-time metrics: because of  this, it can be pulled back to a closed  form on the moduli space
of Riemann surfaces\cite{Becchi:1995ik}. The  component of this form of the appropriate fermionic number can be integrated  on the moduli space of genus $g$ surfaces:  this operation defines topological string amplitudes of the corresponding genus. 

Although the functional (\ref{BRSTPartition}) which defines the string amplitudes does {\it in general}  depend on the superghost field $\gamma^\alpha$,  there are topological models for which one can choose gauge fermions $\Psi[\Phi, g_{\alpha\beta}]$  {\it invariant} under $G_\gamma$
\bea
G_\gamma\int \Psi[\Phi, g_{\alpha\beta}]=0
\label{standardpsi}
\eea
Whenever such a  choice of the gauge fermion is possible,  the  last term in the  action (\ref{mattgaction})  vanishes, the resulting topological string action does not depend on the $\gamma^\alpha$ superghost   and  the coupling   to topological gravity only occurs via the super-current:
\bea
&&\Gamma^{mat+t.g.} [\Phi, g_{\alpha\beta}, \psi_{\alpha\beta}, \gamma^\alpha] =  \Gamma^{matter} [\Phi, g_{\alpha\beta}]  + \int \psi^{\alpha\beta} \, S_{\alpha\beta}[ \Phi, g_{\alpha\beta}] \label{mattgactionN2}
\eea
This is in precisely the situation mostly considered in the literature on topological strings and  the reason why, in those contexts,  the superghost $\gamma^\alpha$  is usually neglected and one  does not bother with the equivariant paradigm.
 In this paper we will show that this point of view is however limited and it comes with a price. First of all  it is too restrictive when analysing situations in which the  $G_\gamma$-invariant choice (\ref{standardpsi}) for the gauge fermion  is not allowed.  We will
elaborate on a specific example when this occurs. 
Moreover, we will explain that  even  in the familiar situation in which  (\ref{standardpsi}) is possible,  neglecting the superghost leads to conceptual puzzles when  one attempts to understand such  an important feature of topological strings as the holomorphic anomaly. 

The typical context in which choice (\ref{standardpsi}) for the gauge fermion is usually possible are the topological matter  theories which are obtained by  twisting  supersymmetric non-linear sigma models  models with {\it extended} $N=2$ supersymmetry \cite{Witten:1988xj}.  
The spinorial supercharges of $N=2$  supersymmetric model transform, upon twisting, into a {\it scalar} nilpotent supercharge which can be identified with the BRST matter operator $S_0^{matter}$ together with a  {\it vector} supercharge
$\hat{G}_\alpha$\footnote{We denote
by $\hat{G}_\alpha$ the vector supercharge of the matter theory, where $\alpha=1,2$ is a vector world-sheet index. This should not generate confusion
with $G_\gamma$, which is the {\it scalar} BRST-operator, where the index $\gamma$ refers to the superghost $\gamma^\alpha$.}. The twisting turns the $N=2$ extended supersymmetry algebra into the topological super-algebra 
\bea
\{S_0, \hat{G}_\alpha\} = P_\alpha\qquad S_0^2=0 = \{ \hat{G}_\alpha, \hat{G}_\beta\}
\label{algebraglobal}
\eea
where $P_\alpha$ are the space-time momentum generators.   Comparing this with (\ref{algebrabrst}),  one is lead to
conjecture that  the equivariant extension $G^{matter}_\gamma$  of the BRST symmetry of twisted $N=2$  matter is obtained
by promoting the  global
vector supersymmetry of the twisted matter model to a {\it local} symmetry
\bea
G^{matter}_\gamma = \gamma^\alpha(x)\, \hat{G}_\alpha
\eea
A legitimate  action for the topological model is the action of the supersymmetric model, appropriately twisted:  this  is invariant under {\it both} the scalar $S_0$ supersymmetry  and the global vector $\hat{G}_\alpha$ supersymmetry:
\bea
\hat{G}_\alpha \Gamma^{matter} [\Phi, g_{\alpha\beta}] = S^{matter}_0 \, \Gamma^{matter}  [\Phi, g_{\alpha\beta}] =0
\eea
Therefore, Noether theorem ensures that, for $\gamma^\alpha(x)$ local
\bea
G^{matter}_\gamma \Gamma^{matter} [\Phi, g_{\alpha\beta}] =- \int  D^\alpha\, \gamma^\beta \, \tilde{S}_{\alpha\beta}
\eea
where $\tilde{S}_{\alpha\beta}$ is the super-current associated to global supercharge $G_\beta$.  It turns out that, in the specific
case of the  twisted $N=2$ supersymmetric non-linear sigma model,   the super-current $\tilde{S}_{\alpha\beta}$ can be taken to be symmetric in the indices $\alpha$ and $\beta$ and    the super-algebra (\ref{algebraglobal}) 
extends to an algebra of local currents
\bea
\{S_0, \tilde{S}_{\alpha\beta}\} = T_{\alpha\beta}
\label{algebralocal}
\eea
where $T_{\alpha\beta}$ is the stress energy tensor. Comparing this with (\ref{supercurrentone}), one identifies $ \tilde{S}_{\alpha\beta}$ with $ S_{\alpha\beta}$, which is therefore conserved
\bea
D^\beta\, S_{\alpha\beta}=0
\eea
In this situation the action one obtains by using  the ``naive'' $S_0$, rather than the
equivariant $S$,
\bea
S_0\, \int \Psi[\Phi, g_{\alpha\beta}]=\Gamma^{matter} [\Phi, g_{\alpha\beta}] + \int \psi^{\alpha\beta} \, S_{\alpha\beta}[ \Phi, g_{\alpha\beta}] 
\eea
is invariant under  both $S_0$ and the {\it local} symmetry $G_\gamma$ which acts on the gravitino  as $G_\gamma \psi^{\alpha\beta} = D^{(\alpha} \gamma^{\beta)}$ (see Eq. (\ref{Sequivariant})). In other words $ \Psi[\Phi, g_{\alpha\beta}]$
can be chosen to be invariant under the local $G_\gamma$ transformations.

Two-dimensional supersymmetric non-linear sigma models enjoy extended $N=2$ supersymmetry when the target space  is  a complex manifold  equipped with
a  K\"ahlerian metric.  One way to twist the  $N=2$  two-dimensional supersymmetric non-linear sigma models leads to the $B$-model \cite{Witten:1988xj}, which is a topological model of the cohomological type.  The physics of the  B-topological  sigma model  is expected to depend on the complex structure of the target manifold but not on the target space metric. In particular  we will see that the B-model can be defined also when the metric on the complex target manifold is not  K\"ahlerian.  In this case the supersymmetric sigma model does not enjoy extended supersymmetry,
the corresponding topological action is {\it not} invariant under the vector supersymmetry, and  topological super-current  $S_{\alpha\beta}$  is not conserved. 
From our discussion above, one does not expect that a choice of a $G_\gamma$-invariant  gauge fermion be possible
in this situation: the coupling to the superghost must necessarily be taken into account for a consistent definition of the string amplitudes.  If one neglects the superghost dependent term in the string action (\ref{mattgaction}) the resulting partition function does not define an equivariant form in the space
of metrics which can be integrated on the moduli space of Riemann surfaces to produce consistent string amplitudes. 
We will see that the equivariant formulation of the B-model coupled to topological gravity 
restores target space metric independence even for complex manifolds for which one cannot pick a K\"ahler metric. 

Even when the $G_\gamma$-invariant choice  for the gauge fermion  is possible
for a {\it fixed} topological matter model, one is often interested in deforming a given matter model and consider the
dependence of the physics on the moduli which parametrize such  deformations.  In the case of the $B$-model among those deformations
are the anti-holomorphic deformations of the complex structure of the target space variety. It turns out that the matter
BRST operator $S_0^{matter}$ is independent of such anti-holomorphic deformations.  Let us denote by $\partial_\abar$ the
anti-holomorphic derivative with respect to the complex  moduli $(m^a, m^\abar)$ which parametrize complex structures of the target space variety. Assume that for a given complex structure of the target space  (\ref{standardpsi}) applies,   so that one can neglect the $\gamma^\alpha$ dependent term in the string action
\bea
&&\Gamma^{mat+t.g.} [\Phi, g_{\alpha\beta}, \psi_{\alpha\beta}, \gamma^\alpha] =  \int S\, \Psi[\Phi, g_{\alpha\beta}]= \int S_0 \Psi[\Phi, g_{\alpha\beta}]
\eea
Taking  the anti-holomorphic derivative $\partial_\abar$ of the string action, one would then obtain
\bea
&& \partial_{\abar}\,\Gamma^{mat+t.g.}= \int \partial_{\abar}\,S_0 \,\Psi[\Phi, g_{\alpha\beta}]= \int S_0 \, \bigl(\partial_{\abar}\, \Psi[\Phi, g_{\alpha\beta}]\bigr)
\label{antiholodef}
\eea
since, as stated above,  $S_0$ is holomorphic in the complex  moduli $(m^a, m^\abar)$. One would then be lead to think that anti-holomorphic deformations  of the target space complex structure are BRST trivial and that  the string amplitudes  are  holomorphic functions of the complex moduli $(m^a, m^\abar)$. Explicit computations show that this  is actually  not the case \cite{Bershadsky:1993ta}:  in the
formulation which neglects $G_\gamma$,  the  non-holomorphicity of the string amplitudes {\it seems} therefore to signal a BRST anomaly.  

This however cannot be the case: a genuine BRST anomaly, like any anomaly of local gauge symmetries,  would  destroy  the consistency
of the corresponding quantum topological string theory. Fortunately for  topological strings, non-holomorphicity of the string amplitudes is in fact not associated
to any anomaly of the {\it equivariant} BRST symmetry ---  which is, as explained,   the relevant notion of BRST symmetry in this context.   To understand this,  consider
the anti-holomorphic derivative of the (\ref{standardpsi})
\bea
0 = \partial_\abar \bigl(G_\gamma  \int  \Psi[\Phi, g_{\alpha\beta}]\bigr)= \int [\partial_\abar, G_\gamma]  \Psi[\Phi, g_{\alpha\beta}]+ \int
G_\gamma \bigl(\partial_\abar \Psi[\Phi, g_{\alpha\beta}]\bigr)
\label{dbarGginvariantpsi}
\eea
It turns out that $G_\gamma$ is {\it not} holomorphic in the complex moduli $(m^a, m^\abar)$, and, correspondingly,  that the deformation of
the gauge fermion $\partial_\abar \Psi[\Phi, g_{\alpha\beta}]$ is not $G_\gamma$-invariant. Therefore the anti-holomorphic deformation in 
(\ref{antiholodef}), although trivial with respect to the ``naive'' $S_0$, is {\it not} trivial with respect to the equivariant $S$. 
Hence anti-holomorphicity is perfectly consistent with BRST invariance with respect to the full, equivariant, $S$. In reality although the equivariant $S$ is not holomorphic,
its anti-holomorphic variation is a $S$-commutator. This ensures that the anti-holomorphicity of the string amplitudes be captured
by  local contact terms which are explicitly calculable.  

The focus of this paper is the relevance of the equivariant superghost of topological gravity to topological strings. It might be useful to add that more  recently it has been understood that the superghost of topological gravity plays a prominent role  also in the topological formulation of localization of supersymmetric quantum field theories in arbitrary dimension \cite{Bae:2015eoa}.

\bigskip

The work contained in this article builds on and extends results obtained years ago with my longtime collaborators, Carlo Becchi and  Stefano Giusto. Those earlier results are contained in Giusto's doctoral dissertation \cite{Giusto:1999}, but were never published. At that time we limited ourselves to considering  B-strings  that are obtainable by twisting 
supersymmetric sigma models  with K\"ahler target manifolds and emphasized the importance of the
superghost of topological gravity for the correct interpretation of the holomorphic anomaly. 
Many years later, Alessandro Tomasiello informed me that he, in collaboration with Anton Kapustin, had considered B-models with non-K\"ahlerian target spaces and had attempted to build topological string models based on them.    Tomasiello and Kapustin were able to define the topological matter model by making use of a target space connection built with the aid of a hermitian but non-K\"ahlerian metric, a construction which   I review in Section \ref{sec:matterB}. As they realized, however, the non-conservation of the super-current  did not allow for a consistent definition of topological string amplitudes.  This result provided me with the motivation for returning to the unpublished work from
my collaboration with Becchi and Giusto and applying  it to the non-K\"ahler situation in order to show that the difficulty encountered  by Kapustin and Tomasiello could be solved by taking into account  the coupling of the matter B-model to the equivariant superghost of topological gravity. As explained in the introduction the coupling to the superghost is, to a certain extent, gauge-dependent.  In order to make my analysis and considerations completely  gauge-independent, and thus more widely applicable,  I decided to extend the work in \cite{Giusto:1999} to the more general BV framework (Section \ref{sec:BV}). In the last section of the present paper, I also generalize the discussion of the holomorphic anomaly found in  \cite{Giusto:1999}, extending it  to B-strings on non-K\"ahler manifolds in a completely gauge-independent set up.

\newsection{BRST transformations of the matter B-model}
\label{sec:matterB}
The topological matter B-model is defined on a complex variety whose complex coordinates we will denote by $(\phi^i\,\phi^\ibar)$. 
The  BRST nilpotent transformation rules  are\footnote{$\rho^i = \rho^i_\alpha\,dx^\alpha$ is a 1-form of ghost number -1, $F^i = \frac{1}{2}\,F^i_{\alpha\beta}\,dx^\alpha\, dx^\beta$, a 2-form of ghost number -2. The sum of form degree and ghost number defines the total fermionic number. Both the BRST operator and the exterior differential are {\it odd} with respect to total fermionic number. }
\bea
&&S_0 \, \phi^i=0\nn\\
&& S_0\, \rho^i = - d\,\phi^i\nn\\
&& S_0 \, F^i = - D\,\rho^i + \frac{1}{2}\, R^i_{\ibar\,j;k}\sigma^\ibar\,\rho^j\,\rho^k\qquad D\,\rho^i\equiv d\rho^i +\Gamma^i_{jk}\, d\phi^j\, \rho^k\nn\\
&&S_0\, \phi^\ibar = \sigma^\ibar\nn\\
&&S_0\, \sigma^\ibar  = 0 
\label{brstmatter}
\eea
In order to preserve covariance of the model under holomorphic reparametrizations of the target space coordinates we introduced a hermitian,  but {\it not necessarily K\"ahler}, metric $g_{i\jbar}$ and a connection
whose non-vanishing components are  $\Gamma^i_{jk}$ and $\Gamma^\ibar_{\jbar \kbar}$\footnote{As I recalled in the
Introduction, this specific connection has been suggested to me by A. Tomasiello.}:
\bea
&&\Gamma^i_{jk} = \frac{1}{2} \, g^{i\,\ibar}\,\bigl( \partial_j\, g_{k\,\ibar}+\partial_k\, g_{j\,\ibar}\bigr)\equiv  g^{i\,\ibar}\,
\partial_{(j }\,  g_{k)\,\ibar}\nn\\
&&\Gamma^\ibar_{\jbar\kbar} = \frac{1}{2} \, g^{\ibar i}\,\bigl( \partial_\jbar\, g_{i\kbar}+\partial_\kbar\, g_{\jbar\,i}\bigr)\equiv  g^{\ibar\,i}\,
\partial_{(\jbar }\,  g_{\kbar)\,i}
\label{NKconnection}
\eea
The non-vanishing components of the curvature are 
\bea
R_{\ibar j;k}^i= \partial_\ibar\,\Gamma^i_{jk}\qquad R^i_{jk;l} = \partial_j\, \Gamma_{kl}^i- \partial_k\, \Gamma_{jl}^i+\Gamma^i_{j m}\, \Gamma^m_{k l}-\Gamma^i_{k m}\, \Gamma^m_{j l}
\eea
and their complex conjugates. If $g_{i\jbar}$ is not K\"ahler, it is not covariantly constant  
\bea
D_i\,g_{j\,\kbar} = \frac{1}{2}\,\bigl( \partial_i\, g_{j\,\kbar}-\partial_j\, g_{i\,\kbar}\bigr)\equiv \partial_{[i}\,g_{j],\kbar}\equiv C_{[ij];\kbar}
\eea
The action is a pure gauge
\bea
\Gamma_0 =\int  S_0 \,\Psi
\eea
where $\Psi$ is the gauge fermion.  The traditional choice is
\bea
\Psi = \theta_i\, F^i + g_{i \jbar} \, \rho^i \star\, d\,\phi^\jbar
\label{gferm}
\eea
where one has introduced a  trivial BRST doublet 
\bea
&&S_0 \, \theta_i = H_i\qquad S_0\,H_i =0
\label{trivialTheta}
\eea
$\theta_i$ has ghost number $+1$ and plays the function of  Nakanishi-Laudrup field, $H_i$ is a Lagrangian multiplier. The gauge fermion (\ref{gferm})
gives  non-degenerate kinetic terms for all fields:
\bea
&&\!\!\!\!\Gamma_0 = H_i\, F^i - \theta_i \,  D\,\rho^i - \frac{1}{2}\, \theta_i\, R^i_{\ibar\,j;k}\sigma^\ibar\,\rho^j\,\rho^k+ g_{i\ibar}\, d\phi^i \star\,d\, \phi^\ibar -  g_{i \jbar} \, \rho^i \star\, d\,\sigma^\jbar- \rho^i\star\, d\phi^\jbar\,\sigma^\kbar \partial_\kbar\,g_{i\jbar}=\nn\\
&&\qquad = H_i\, F^i - \theta_i \,  D\,\rho^i - \frac{1}{2}\, \theta_i\, R^i_{\ibar\,j;k}\sigma^\ibar\,\rho^j\,\rho^k+ g_{i\ibar}\, d\phi^i \star\,d\, \phi^\ibar+\nn\\
&&\qquad\qquad-  g_{i \jbar} \, \rho^i \star\, D\,\sigma^\jbar- \rho^i\star\, d\phi^\jbar\,\sigma^\kbar\, C_{[\kbar\jbar]; i}
\label{matteraction}
\eea
In the K\"ahler case this action is obtained by twisting of the action of the $N=(2,2)$ supersymmetric non-linear
sigma model.

\newsection{The coupling  to topological gravity}
The coupling of the B-model to topological gravity is determined by requiring  the validity of the super-algebra (\ref{algebrabrst}). 
For example, applying  both two sides of Eq. (\ref{algebrabrst}) to $\phi^i$ one obtains
 \bea
&&\ {\cal L}_\gamma \phi^i = i_\gamma (d\,\phi^i) = \{S_0, G_\gamma\}\, \phi^i = S_0\,\bigl( G_\gamma(\phi^i)\bigr)= -i_\gamma (S_0\, \rho^i) = S_0\, i_\gamma(\rho^i)
\eea
where we introduced $i_\gamma$, the operation which contract a form along the superghost vector field $\gamma^\alpha$, and used the fact that
\bea
\mathcal{L}_\gamma = \{ d, i_\gamma\}
\eea
on forms. One thus derives
\bea
G_\gamma(\phi^i) =  i_\gamma(\rho^i)
\eea
Proceeding in this way one obtains the following BRST transformations
\bea
&&\hat{S}\, F^i \equiv S\, F^i + \Gamma^i_{i_\gamma(\rho) j}\,F^j= -D\,\rho^i +\frac{1}{3}\, R^i _{k j;l}\,i_\gamma(\rho^k)\, \rho^j\, \rho^l + \frac{1}{2}\, R^i _{\kbar j;l}\, \sigma^\kbar \rho^j\, \rho^l\nn\\
&& \hat{S}\,\rho^i  \equiv S\, \rho^i + \Gamma^i_{i_\gamma(\rho) j}\,\rho^j= -d\,\phi^i +  i_\gamma(F^i) \nn\\
&&  S\, \phi^i =i_\gamma(\rho^i)\nn\\
&&S \, \phi^\ibar =\sigma^\ibar\nn\\
&&  \hat{S}\,\sigma^\ibar \equiv S\, \sigma^\ibar + \Gamma^\ibar_{\sigma \jbar}\,\sigma^\jbar =i_\gamma\, d\,\phi^\ibar\nn\\
&& \hat{S}\, \theta_i \equiv S\, \theta_i -\Gamma^j_{i_\gamma(\rho) i}\,\theta_j=  H_i\nn\\
&& \hat{S}\, H_i \equiv S\, H_i -\Gamma^j_{i_\gamma(\rho) i}\,H_j=i_\gamma\, D\,\theta_i - R_{\kbar l i}^j \,\sigma^\kbar\, i_\gamma(\rho^l)\,\theta_j-\frac{1}{2}\,R_{k l i}^j \,i_\gamma(\rho^k)\, i_\gamma(\rho^l)\,\theta_j
\label{Scurvedone}
\eea 
To simplify computations and notation we introduced, for all fields but the coordinate fields $(\phi^i, \phi^\ibar)$,
the BRST operator $\hat{S}$,  covariant under target space holomorphic reparametrizations
\bea
\hat{S} \equiv S + \Gamma_{i_\gamma(\rho)}\oplus \bar{\Gamma}_{\sigma}
\eea
where the connection pieces are matrices acting on (anti-)holomorphic indices in the standard way. 
$\hat{S}$ acts on the coordinates fields  via  covariant connections
\bea
\hat{S} = \sigma^\ibar \, D_\ibar + i_\gamma(\rho^i) \, D_i +\cdots
\eea
Therefore
\bea
\hat{S} \Psi =  S \Psi
\eea
on functionals $\Psi$ which are invariant under target space holomorphic reparametrizations.

The
relation $S^2 ={\cal L}_\gamma$ translates  into the following relation for $\hat{S}$: 
\bea
&&\hat{S}^2 = \{i_\gamma, D\} + \Bigl(\frac{1}{2}\,R_{i_\gamma(\rho)\, i_\gamma(\rho)}+R_{\sigma\, i_\gamma(\rho)}\Bigr)\oplus  \Bigl(\frac{1}{2}\,\bar{R}_{\sigma\, \sigma}+\bar{R}_{\sigma\, i_\gamma(\rho)}\Bigr)=\nn\\
&&\qquad \equiv \{i_\gamma, D\} +\frac{1}{2}\,R_{\chi\,\chi}\oplus \frac{1}{2}\,\bar{R}_{\chi\,\chi}
\label{hatSsquared}
\eea
where 
\bea
&&R_{i_\gamma(\rho)\, i_\gamma(\rho)} \equiv  R_{ij}\,i_\gamma(\rho^i)\,i_\gamma(\rho^j)\nn\\
&&R_{\sigma, i_\gamma(\rho)}\equiv R_{\ibar i}\, \sigma^\ibar\,i_\gamma(\rho^i)\nn\\
&&\bar{R}_{\sigma\,\sigma} \equiv \bar{R}_{\ibar \jbar}\, \sigma^\ibar\,\sigma^\jbar\nn\\
&&\bar{R}_{i_\gamma(\rho)\,\sigma}\equiv \bar{R}_{i\jbar}\,i_\gamma(\rho^i)\,\sigma^\jbar\nn\\
&&R_{\chi\,\chi} \equiv R_{i_\gamma(\rho)\, i_\gamma(\rho)}+R_{\sigma\, i_\gamma(\rho)}+R_{i_\gamma(\rho)\,\sigma}
\eea
are the matrix-valued curvature 2-forms acting on (anti-)holomorphic indices in the standard way,
\bea
D  = d + \Gamma_{d\phi}\oplus \bar{\Gamma}_{d\bar{\phi}}
\eea
is the covariant derivative of ghost number 0, and 
\bea
\chi^I\equiv \bigl(i_\gamma(\rho^i), \sigma^\ibar\bigr)\qquad I=(i,\ibar)
\eea
is a world-sheet scalar of ghost number 1 which acts like the 1-differential on the target space.

\newsection{The gauge fermion}

The action of the coupled model is 
\bea
\Gamma =\int  \bigl(S_0 + G_\gamma\bigr)\, \Psi = \Gamma^{matter}_0 +  \int \psi_{\alpha\beta}\, \frac{\delta \Psi}{\delta\, g_{\alpha\beta}} + \int G_\gamma\, \Psi
\eea
As recalled in the Introduction, the first term is the ``matter'' action. The second term is the standard coupling of the topological gravitino to the matter super-current
\bea
S_{\alpha\beta} =  \frac{\delta \Psi}{\delta\, g_{\alpha\beta}}
\label{mattersupercurrent}
\eea
which produces the insertions in  topological string amplitudes which are analogous to the $b$-zero modes insertions of bosonic
string theory.

The third  term in the action does not usually appear in the ``naive'' recipe for coupling topological matter to topological
gravity.   With the usual choice for the gauge fermion (\ref{gferm}),  this term is 
\bea
&&G_\gamma\, \Psi =-\theta_i\,  \frac{1}{3}\, R^i_{jkl}\, i_\gamma(\rho^j) \,\rho^k\, \rho^l +C_{ij}^k\, i_\gamma(\rho^i)\, g_{k\kbar}\,\rho^j\star d\phi^\kbar + g_{i\,\jbar}\, i_\gamma(F^i)\,\star\,d\phi^\jbar
\label{GgammanonK}
\eea
As a matter of fact, this  does not vanish even when the target space metric is  K\"ahler, in which case it reduces to
\bea
&&G_\gamma\, \Psi = g_{i\,\jbar}\, i_\gamma(F^i)\,\star\,d\phi^\jbar=  F^i \ (  g_{i\,\jbar}\, i_\gamma(\,\star\,d\phi^\jbar))
\label{KGgamma}
\eea
Recalling the matter action (\ref{matteraction}) one sees however that  all the $\gamma^\alpha$ dependence of the 
string action is confined, in the   K\"ahler case, to the auxiliary sector $(F^i, H_i)$
\bea
\Gamma =  F^i\,( H_i  +  g_{i\,\jbar}\, i_\gamma(\,\star\,d\phi^\jbar)) +\cdots
\eea
Hence the field redefinition 
\bea
\tilde{H}_i =  H_i  +  g_{i\,\jbar}\, i_\gamma(\,\star\,d\phi^\jbar)
\label{Hredefinition}
\eea
is sufficient to eliminate the term linear in the auxiliary field $F^i$ in $G_\gamma\Psi$ and  the $\gamma^\alpha$-dependence  altogether from the string action when the metric is K\"ahler.  Note that redefinition (\ref{Hredefinition}) of the Lagrangian multiplier $H_i$ amount to modifying the BRST transformations of the Nakanishi-Laudrup field $\theta_i$
 \bea
 \hat{S}\, \theta_i =  \tilde{H}_i -g_{i\,\jbar}\, i_\gamma(\,\star\,d\phi^\jbar)
 \eea
 by a term which depends on the world-sheet metric, i.e. a term  which is  {\it not} topological. This is harmless for the topological character of the theory since this is confined to the BRST- trivial $(\theta_i, H_i)$ sector.

When the metric is not K\"ahlerian,  however, even  after the redefinition (\ref{Hredefinition}), one remains with non-vanishing coupling to the superghost
\bea
&&G_\gamma\, \Psi =-\theta_i\,  \frac{1}{3}\, R^i_{jkl}\, i_\gamma(\rho^j) \,\rho^k\, \rho^l +C_{ij}^k\, i_\gamma(\rho^i)\, g_{k\kbar}\,\rho^j\star d\phi^\kbar 
\label{nonequigf}
\eea
In this case the consistent coupling of topological matter to topological gravity cannot be described only by the standard interaction with the gravitino ---  it requires superghost dependent terms. The reason for this is that when the
target space metric is not K\"ahler, the corresponding  supersymmetric model does not enjoy extended supersymmetry and the super-current $S_{\alpha\beta}$ is not conserved.

To check this, let us remark that in the $G_\gamma$ transformation laws  of the matter sector, $\gamma^\alpha$ appears with no derivatives, so we can write\footnote{We denote
by $\hat{G}_\alpha$ the vector supercharges of the matter theory, where $\alpha=1,2$ is a vector world-sheet index. This should not generate confusion
with $G_\gamma$, which is the {\it scalar} BRST-operator, where the index $\gamma$ refers to the superghost $\gamma^\alpha$.}
 \bea
G_\gamma = \gamma^\alpha\, \hat{G}_\alpha \qquad  \text{in the matter sector}
\eea
where $\hat{G}_\alpha$ are the vector super-charges of the matter theory satisfying 
\bea
\{S^{matter}_0, \hat{G}_\alpha\} = P_\alpha \qquad  \text{in the matter sector}
\eea
Hence
\bea
\hat{G}_\alpha\, \Gamma_0 = - S_0\, \int \hat{G}_\alpha\, \Psi
\label{matterGinvariance}
\eea
We see therefore that $G_\gamma$-invariance of $\Psi$ ensures conservation of the super-current $S_{\alpha\beta}$. Conversely,  if $\Psi$ is not $\hat{G}_\alpha$-invariant, we obtain, via the Noether procedure,  the (non-)conservation equation for the super-current
\bea
D^\alpha \, S_{\alpha\beta} = S_0 \, \hat{G}_\beta\,\Psi = S_0\, \bigl(C_{ij}^k\,  g_{k\kbar}\,\rho^i_\alpha\,g^{\beta\gamma}\rho^j_\beta\,\partial_\gamma\phi^\kbar- \frac{1}{3}\, \theta_i\, R^i_{jk; l}\,\rho^j_\alpha \,\epsilon^{\beta\gamma}\,\rho^k_\beta\, \rho^l_\gamma )
\eea
which does not vanish in the non-K\"ahlerian case.

\newsection{BV formulation}
\label{sec:BV}
The action of a topological model of the cohomological type, like the B-model,  is BRST-trivial\footnote{In recent times, theories of this type are being called ``localizable''.}
\bea
\Gamma = \int S\,\Psi
\eea
The reason is that the {\it local} BRST cohomology of such class of models is empty --- therefore no non-trivial invariant
term can show up in the action. 

All the parameters, or coupling constants, which appear in the gauge fermion $\Psi$ are gauge parameters, and the physics
does not depend on them. The physical parameters of cohomological models are contained in the BRST operator itself. We have seen that,  in the case of the B-model, in order to define $S$ we had to specify not only a complex structure on the target space, but  also ---
to preserve target space holomorphic reparametrization invariance --- 
a metric on it. Therefore, in principle, the physics could depend on both.

Let us denote by $\delta$ a generic deformation of the parameters  --- complex structure and metric --- on which $S$ depends. The
variation of the action under such a deformation takes the form
\bea
\delta \,\Gamma = \int \delta(S)\, \Psi + S\, \delta(\Psi)
\eea
Let us denote by 
\bea
I \equiv \delta(S)
\label{Sdeformation}
\eea
the operator of ghost number +1 which is obtained by deforming $S$. Since 
\bea
S^2 = \mathcal{L}_\gamma
\eea
we obtain
\bea
\{I, S\}=0
\label{Ideformation}
\eea
Hence the deformation of the action satisfies
\bea
S\, (\delta \,\Gamma) = \int  S\, I\, \Psi  = - I\,  \int S\, \Psi =- I\,\Gamma
\eea
i.e.  the deformation of the action is BRST-closed {\it modulo the equations of motion} generated by $I$. 
Deformations $I$ which are $S$-commutators 
\bea
I = [S, L]
\eea
with $L$ bosonic, must be considered trivial solutions of the consistency equation (\ref{Ideformation}). Indeed, in this case the corresponding deformation of the action
\bea
\delta \,\Gamma = S\, \int (L \Psi + \delta\,\Psi) - L \, \Gamma
\eea
is BRST-trivial modulo the equations of motion generated by $L$.

This is the general paradigm of cohomological theories: Since the local BRST cohomology is empty, cohomological
theories have no standard --- i.e. BRST invariant --- observables. Their only local observables are contained in the local BRST cohomology
modulo the equation of motions.  Because of this, a gauge-independent analysis of a topological theory of the cohomological type 
requires upgrading the usual BRST  framework to the Batalin-Vilkovisky  one. 

The BV action corresponding to the BRST transformations (\ref{Scurvedone}) is
\bea
&& \Gamma_{BV} = i_\gamma(\rho^i)\, \phi_i^* + (-d\phi^i + i_\gamma(F^i) - \Gamma_{jk}^i\, i_\gamma(\rho^j)\, \rho^k)\, \rho_i^* +\nn\\
&&\qquad +  (  -D\,\rho^i +\frac{1}{3}\, R^i _{k j;l}\,i_\gamma(\rho^k)\, \rho^j\, \rho^l + \frac{1}{2}\, R^i _{\kbar j;l}\, \sigma^\kbar \rho^j\, \rho^l - \Gamma^i_{jk}\, i_\gamma(\rho^j)\, F^k)\, F_i^*+\nn\\
&& \qquad +\sigma^\ibar\, \phi_\ibar^* + i_\gamma(d\phi^\ibar)\, \sigma^*_\ibar+\nn\\
&&\qquad \qquad\qquad+\psi_{\alpha\beta}\,g^{*\;\alpha\beta}+{\cal L}_\gamma\,g_{\alpha\beta}
\,\psi^{*\;\alpha\beta}
\eea
We introduced the anti-fields, denoted with the asterisk, in correspondence of each of the fields of the model, but, for simplicity,  we neglected the trivial $(\theta_i, H^i)$ doublet which plays no role in the gauge-independent physics of the model.

From this  action one obtains the BRST transformations for the anti-fields
\bea
&&\hat{S}\, F^{*}_i = i_\gamma(\rho_i^*)  \nn\\
&& \hat{S}\,\rho^*_i = -\bigl( D \, F^*_i +(R^j_{\sigma \,\rho\; i}+ \frac{1}{2}\,R^j_{i_\gamma(\rho)\,\rho\; i})\,F_j^*\bigr) +i_\gamma(\tilde{\phi}_i^*) \nn\\
&& \hat{S}\, \tilde{\phi}^*_i =- D\,\rho_i^*- \bigl(R^j_{\sigma \rho\; i}+ \frac{1}{2}\,R^j_{i_\gamma(\rho) \rho\; i}\bigr)\,\rho_j^* +\nn\\
&&\qquad  -  \bigl(R^j_{\sigma F\; i}+ \frac{1}{2}\,R^j_{i_\gamma(\rho)\, F\; i}+ R^j_{d\bar{\phi}\,\rho\; i}+R^j_{d\phi\,\rho\; i}+\frac{1}{2}\,D_\rho R_{\sigma\,\rho\; i}^j+\frac{1}{6}\,D_\rho\, R_{i_\gamma(\rho)\,\rho\;i}^j\bigr)\, F^*_j
\label{antiBRSTcurvedone}
\eea
where $\tilde{\phi}^*_i$ is related to the anti-field $\phi_i^*$ by the formula
\bea
\tilde{\phi}^*_i = \phi_i^*+\frac{1}{2}\,R^k _{ij;l}\,\rho^j\, \rho^l \, F^*_k-\Gamma^k_{ij}\,F^j\, F_k^* -\Gamma^k_{ij}\,\rho^j\, \rho_k^* 
\eea

\newsection{Observables}
\hypersetup{pageanchor=false}

In the BV formalism,  the observables --- which,  for the B-model, as explained,  are BRST classes on the space of fields modulo the equations of motion ---  map to BRST cohomology classes on the space of both fields and antifields.  

Local observables satisfy BRST descent equations. In the equivariant context, these are descent equations which involve the BRST operator $S$, the exterior differential on forms $d$ and $i_\gamma$ the contraction of forms along the vector field $\gamma^\alpha$. 
They take the form
\bea
&& S\, O^{(0)}(x) = i_\gamma(O^{(1)}(x))\nn\\
&& S\, O^{(1)}(x) +d\, O^{(0)}= i_\gamma(O^{(2)}(x))\nn\\
&& S\, O^{(2)}(x) +d\, O^{(1)}=0
\label{equivariantdescent}
\eea
where $O^{(0)}(x), O^{(1)}(x)$ are the ``descendant'' of the observable 2-form $ O^{(2)}(x)$ whose integral is BRST invariant.

Descent equations can be nicely written in terms of the nilpotent coboundary operator $\delta$
\bea
\delta \equiv S + d - i_\gamma
\eea
acting on polyforms, which are sum of forms of different degrees. Indeed (\ref{equivariantdescent}) can be recast as
\bea
\delta O(x) = 0\qquad O(x)= O^{(0)}(x)+O^{(1)}(x)+O^{(2)}(x)
\eea
Since our observables will necessarily contain the anti-fields $\phi_i^*, \rho_i^*, F_i^*$ which make up a polyform
with values in the holomorphic cotangent of the target space, it will be useful 
to derive the analogs of (\ref{equivariantdescent}) for polyforms with values in the holomorphic tangent:
\bea
O^i(x)= O^{i\;(0)}(x)+O^{i\;(1)}(x)+O^{i\;(2)}(x)
\eea
Starting from
\bea
\hat{S}\, O^{i\;(0)}(x)= i_\gamma(O^{i\;(1)}(x))
\eea
and using the algebra (\ref{hatSsquared}) for the covariant BRST operator $\hat{S}$ one derives:
\bea
&& \hat{S}\, O^{i\;(0)}= i_\gamma( O^{i\;(1)})\nn\\
&& \hat{S}\, O^{i\;(1)}+ D\,O^{i\; (0)} +\frac{1}{2} \,\mathcal{R}^i_j\, O^{j\; (0)}  = i_\gamma(O^{i\;(2)})\nn\\
&&  \hat{S}\,O^{i\;(2)}+D\, O^{i\;(1)} +\frac{1}{2} \,\mathcal{R}^i_j\,O^{(1)}+ \, \Bigl[\frac{1}{2}\,R_{F\, i_\gamma(\rho)} + R_{F\, \sigma}+\frac{1}{2}\,R_{\rho\,d\phi}+   R_{\rho\, d\,\bar{\phi}}+ \nn\\
 &&\qquad + \frac{1}{2}\, D_\rho\,R_{\rho\,\sigma}+\frac{1}{6}\,D_\rho\, R_{\rho\,i_\gamma(\rho)}
\Bigr]^i_j\, O^{j\;(0)}=0
\label{covariantdescent}
\eea
where we introduced the matrix-valued two-form
\bea
&&\mathcal{R}^i_j \equiv (R_{\rho\, i_\gamma(\rho)})^i_j+ 2\, (R_{\rho\, \sigma})^i_j
\label{descentthreetris}
  \eea
Comparing (\ref{covariantdescent}) with the BRST transformation rules of the anti-fields, we see that  $F^*_i, \rho_i^*$ and $\tilde{\phi}_i^*$\footnote{Note that it is $\tilde{\phi}^*_i$ and not the anti-field $\phi_i^*$ which satisfies the descent together with $F^*_i$ and $\rho_i^*$.}  satisfy  descent equations which are the analogs of  (\ref{covariantdescent}) for operators which are valued in the holomorphic cotangent
bundle.

 \subsection{The  differential geometry of complex structure moduli space}
  \label{subsec:diffgeo}
  The dependence of the action on the complex structure is parametrized by Beltrami differentials $\mu^i_\jbar$ 
  \bea
  d\phi^i = \Lambda^i_j \, \bigl(d\phi_0^j + \mu^j_\jbar\, d\phi_0^\jbar\bigr)
\label{beltramione}
  \eea
 where $(\phi_0^i, \phi^\ibar_0)$ is a {\it fixed} system of complex coordinates, and $\Lambda_j^i$ are the integrating factors,
which are (non-local) functionals of the Beltrami differentials $\mu^i_\jbar$.  Derivative with respect to the complex
structure moduli is performed keeping   $(\phi_0^i, \phi^\ibar_0)$ fixed.  Let  $\partial_a$ denote the holomorphic derivative with respect to
the complex structure moduli coordinates $\{m^a\}$. The Beltrami differentials $\mu^i_\jbar$ are holomorphic
functions of the moduli coordinates $m^a$
\bea
&& \partial_a\, d\phi^i= \partial_a \Lambda^i_j \,(\Lambda^{-1})^j_{k}\, d\phi^k +  \Lambda^i_j \, \partial_a \mu^i_\jbar\, d\phi^\jbar \equiv  A^i_{a\;j}\, d\phi^j + \mu^i_{a\; \jbar}\, d\phi^\jbar
\label{beltramitwo}
\eea 
where
\bea
&&A^i_{a\;j} = \partial_j\bigl[\partial_a\,\phi^i(\phi_0,m)\bigr]_{\phi_0=\phi_0(\phi, m)}\qquad \mu^i_{a\;\jbar} = \partial_\jbar\bigl[\partial_a\,\phi^i(\phi_0,m)\bigr]_{\phi_0=\phi_0(\phi, m)}
\eea
$ A^i_{a\;j}$ transforms as a connection under $m^a$-dependent holomorphic reparametrizations of the target space coordinates
$\phi^i$ while 
\bea
 \mu^i_a \equiv  \mu^i_{a\; \jbar}\, d\phi^\jbar
 \eea  
are $(0,1)$-forms with values in the holomorphic tangent which are {\it closed} under the Dolbeault exterior differential in the $(\phi^i, \phi^\ibar)$ complex structure
\bea
\partialbar \mu^i_a =0 \qquad 
\label{ksone}
\eea
Moreover
\bea
&& \partial_{[j}\,A^i_{k]}=0\qquad \partial_{\jbar}\,A^i_k=\partial_k\, \mu^i_\jbar 
\eea
These equations are equivalent to the Kodaira-Spencer equation for the Beltrami differential in the fixed
system of complex coordinates $(\phi^i_0,\phi_0^\ibar)$:
\bea
\partialbar \mu^i - \mu^j \,\partial_j\mu^i =0\qquad \mu^i\equiv \mu^i_\jbar\,d\phi_0^\jbar
\eea
Eq. (\ref{beltramitwo}) leads to the definition of a {\it covariant} holomorphic derivative with respect to
the complex structure moduli
\bea
D_a \, d\phi^i\equiv  \partial_a\, d\phi^i-A^i_{a\;j}\, d\phi^j =\mu^i_{a}
\label{mderholodiff}
\eea
The relations dual to (\ref{mderholodiff}) which capture the moduli dependence of the holomorphic
derivatives are
\bea
\bigl[ \partial_a, \partial_\ibar\bigr]= -\mu^i_{a\; \ibar}\,\partial_i\qquad \bigl[ \partial_a, \partial_i\bigr]= -A^j_{a\; i}\,\partial_j
\eea
For example, on a holomorphic vector $V^i$ one has
\bea
&&  \bigl[ D_a, \partial_i\bigr] V^j= \partial_i A^j_{a\; k}\, V^k\qquad \bigl[ D_a, \partial_\ibar\bigr]\, V^j=-\mu^i_{a\; \ibar}\,\partial_i\, V^j+ \partial_k \mu^j_{a\;\ibar}\,V^k
\eea
We will also be interested in evaluating the moduli dependence of target space covariant derivatives built with a
connection $\Gamma^i_{jk}$ like (\ref{NKconnection}).  
\bea
&& [D_a, D_k] = (\mathcal{H}_{a k})\qquad [D_a, D_\kbar] = - \mu^k_{a\;\kbar} \, D_k + (\mathcal{H}_{a\kbar})
\label{DDcommutators}
\eea
where $(\mathcal{H}_{a k})$ and $(\mathcal{H}_{a \kbar})$ are matrices acting
on the (anti-)holomorphic tangent indices. For example when acting on holomorphic vectors one finds that
\bea
\bigl(\mathcal{H}_{a \kbar}\bigr)^i_j=  D_j\,\mu_{a\kbar}^i
\label{Hholoone}
\eea
Moreover, it turns out that 
\bea
&&  \bigl(\mathcal{H}_{a k}\bigr)^i_j = \partial_k\,(A_a)^i_j+D_a \Gamma^i_{kj}=\delta_a\, \Gamma^i_{jk}= D_a\, g^{i\jbar}\, \partial_{(k}\, g_{j)\jbar}+  g^{i\jbar}\, \partial_{(k} D_{a}\, g_{j)\jbar}
\eea
where $\delta_a \Gamma^i_{jk}$ is the variation of the connection $\Gamma^i_{jk}$  induced by the variation of the metric  $g_{i\jbar}$ 
\bea
\delta_a\, g_{i\jbar} = D_a \, g_{i\jbar}
\label{holometricvar}
\eea
which can accompany a  complex structure deformation. This variation is {\it independent} of the variation of the complex structure
and it is {\it arbitrary}.  We can recast (\ref{DDcommutators}), when acting on holomorphic vector fields,  as
\bea
&&  [D_a, D] = \delta_a\, \Gamma_{d\phi} + d\phi^\kbar\,( D_j\,\mu_{a\kbar}^i)
\label{holoconnection}
\eea
where $D$ is the covariant exterior differential 
\bea
DV^k= dV^k+ d\phi^i\,\Gamma^k_{i j}\,V^j
\eea
We can analogously compute the dependence of the curvature built with $\Gamma^i_{jk}$ on the complex structure moduli:
\bea
&& D_a\, R^i_{\ibar j;k}= \delta_a \, R^i_{\ibar j;k}-\frac{1}{2}\,\{D_k ,D_j\} \,\mu^{i}_{a \ibar}+\frac{1}{2}\, (R^i_{kl;j}+R^i_{jl;k})\,\mu^l_{a\;k}\nn\\
&& D_a\, R^i_{ jk;l}= \delta_a\, R^i_{ jk;l}= D_{[j}\, \delta_a\,\Gamma^i_{k]l}
\label{holocurvatures}
\eea
Again here we denoted with $ \delta_a \, R^i_{\ibar j;k}$ and $\delta_a\, R^i_{ jk;l}$ the variation of the curvatures induced by the
variation (\ref{holometricvar}) of the metric  $g_{i\jbar}$.

We will be also interested in taking the anti-holomorphic derivatives of the coordinate fields with respect to anti-holomorphic
moduli $m^\abar$, which we will denote by $\partial_\abar$. The anti-holomorphic Beltrami differentials are defined in a way
analogous to (\ref{beltramione})
\bea
 d\phi^\ibar = \bar{\Lambda}^\ibar_\jbar \, \bigl(d\phi_0^\jbar + \mu^\jbar_ j\, d\phi_0^j\bigr)
\label{beltramioneanti}
  \eea
The anti-holomorphic derivative $\partial_\abar$ acts in a way analogous to (\ref{beltramitwo}):
\bea
&& \partial_\abar\, d\phi^\ibar= \partial_\abar \Lambda^\ibar_\jbar \,(\Lambda^{-1})^\jbar_{\kbar}\, d\phi^\kbar +  \Lambda^\ibar_\jbar \, \partial_\abar \mu^\ibar_j\, d\phi^j \equiv  A^\ibar_{\abar\;\jbar}\, d\phi^\jbar + \mubar^\ibar_{\abar\; j}\, d\phi^j
\label{beltramitwoanti}
\eea 
where
\bea
 \mubar^\ibar_\abar \equiv  \mubar^\ibar_{\abar\; j}\, d\phi^j
 \eea  
are $(1,0)$-forms with values in the anti-holomorphic tangent which are {\it closed} under the anti-Dolbeault exterior differential in the $(\phi^i, \phi^\ibar)$ complex structure
\bea
\partial \mubar^\ibar_a =0 \qquad 
\label{ksoneanti}
\eea
The covariant derivative is defined in the same way as (\ref{mderholodiff}) 
\bea
D_\abar \, d\phi^\ibar \equiv  \partial_\abar\, d\phi^\ibar-A^\ibar_{\abar\;\jbar}\, d\phi^\jbar =\mu^\ibar_{\abar}
\label{mderholodiffanti}
\eea
and it satisfies
\bea
&&  \bigl[ D_\abar, \partial_\ibar \bigr] V^\jbar=-\mubar^\ibar_{\abar\; i}\,\partial_\ibar\, V^j+ \partial_\kbar \mubar^\jbar_{\abar\;i}\,V^k
\eea
The anti-holomorphic moduli dependence of target space covariant derivatives built with a
connection $\Gamma^i_{jk}$ like (\ref{NKconnection}) are captured by 
\bea
&& [D_\abar, D_\kbar] = (\mathcal{H}_{\abar \kbar})\qquad [D_\abar, D_ k] = - \mubar^\kbar_{\abar\;k} \, D_\kbar + (\mathcal{H}_{\abar k})
\label{DDcommutatorsanti}
\eea
where $(\mathcal{H}_{\abar \kbar})$ and $(\mathcal{H}_{\abar k})$ are matrices: when acting
on the anti-holomorphic indices they are given by the complex conjugate of (\ref{Hholoone})
\bea
&&  \bigl(\mathcal{H}_{\abar k}\bigr)^\ibar_\jbar =   D_\jbar\,\mu_{\abar k}^\ibar\qquad   \bigl(\mathcal{H}_{\abar \kbar}\bigr)^\ibar_\jbar= \delta_\abar\, \Gamma^\ibar_{\jbar\kbar}
\label{Hantiholoone}
\eea
where $\delta_\abar \Gamma^\ibar_{\jbar\kbar}$ is the variation of the connection $\Gamma^\ibar_{\jbar\kbar}$  induced by the (arbitrary) variation of the metric  $g_{i\jbar}$ 
\bea
\delta_\abar\, g_{i\jbar} = D_\abar \, g_{i\jbar}
\eea
which can accompany a  complex structure  anti-holomorphic deformation. 
When acting on holomorphic target space indices, the matrices  $(\mathcal{H}_{\abar \kbar})$ and $(\mathcal{H}_{\abar k})$ write instead
\bea
&& (\mathcal{H}_{\abar \kbar})^i_j= 0\qquad (\mathcal{H}_{\abar k})^i_j= D_\abar\, \Gamma^i_{kj}= 
\delta_\abar\Gamma^i_{kj}+ D^i\,\mubar_{\abar\; k j}
\eea
where
\bea
&& \mubar_{\abar\; i j}\equiv \frac{1}{2}\, (\mu^\ibar_{a\,i}\, g_{j\ibar}+\mu^\ibar_{a\,j}\, g_{i\ibar})
\eea
is the anti-Beltrami differential with lower {\it symmetrized} holomorphic indices. Note that in the K\"ahler case $\mubar_{\abar\; i j}=
\mubar_{\abar\;j}^\ibar\, g_{\ibar i}$ is automatically symmetric, but this is not so for the non-K\"ahlerian metric we are considering.
In conclusion the following commutation relation holds  on holomorphic vectors
\bea
&& [D_\abar, D]\,V^i= \delta_a\Gamma^i_{d\phi j}\, V^j+
\,D^i\,\mubar_{\abar\; k j}\,d\phi^k\, V^j
\eea
We can analogously compute the dependence of the curvature on the anti-holomorphic moduli
\bea
&& D_\abar\, R^i_{\ibar j;k}=\delta_\abar R^i_{\jbar j; k}+ D_\ibar \,D^i (\mubar_{\abar\; jk})\nn\\
&& D_\abar\, R^i_{ jk;l}= \delta_\abar\, R^i_{ jk;l} -\mu_{\abar\, [j}^\mbar\,R^i_{\mbar k];l} + D_{[j} \,D^i (\mubar_{\abar\; k]l})
\label{antiholocurvature}
\eea
The tensor 
\bea
&&\mathcal{R}_{\abar\;jk;l}^i\equiv -\mu_{\abar\, [j}^\mbar\,R^i_{\mbar k];l} + D_{[j} \,D^i\,\mubar_{\abar\; k]l}
\label{Rmutensor}
\eea
vanishes when the metric is  K\"ahler, thanks to  the following relations which hold in this case, 
\bea
g^{i\ibar}\, R_{j\ibar ;l}^m=  g^{m\ibar}\, R_{j\ibar;l}^i\qquad \mu_{\abar\; ij} = g_{i\jbar}\, \mu^\jbar_{\abar\;j}\qquad D_{j}\,g^{i\ibar}=0
\label{KahlerRidenities}
\eea 
Finally, the anti-fields $\phi_i^*$ and $\phi_\ibar^*$ transform as the holomorphic derivatives $\partial_i$ and $\partial_\ibar$ and therefore
\bea
&& D_a \phi_i^* =  0\qquad D_a\,\phi^*_\ibar =- \mu^i_{a\;\ibar}\,\phi^*_i\nn\\
&& D_\abar \phi_i^* =- \mu^\ibar_{\abar\;i}\,\phi^*_\ibar\  \qquad D_\abar\,\phi^*_\ibar =0
\label{holoantifields}
\eea

\subsection{Observables associated to complex structure deformations}

To any Beltrami differential $\mu^i_{a\;\jbar}$ we can associate the 0-form of ghost number +1 with values in the holomorphic tangent
\bea
M_a^{i\;(0)}=  \mu^i_{a\; \jbar}\,\sigma^\jbar
\eea
Thanks to the Beltrami equation (\ref{ksone}) this satisfies
\bea
&& \hat{S}\, \bigl( \mu^i_{a\; \jbar}\,\sigma^\jbar\bigr) = i_\gamma\bigl(\mu^i_{a\; \jbar}\,d\,\phi^\jbar+  D_k\, \mu^i_{a\; \jbar}\, \rho^k\, \sigma^\jbar\bigr) 
\eea
We can therefore construct  the corresponding  one-form $M_a^{i\;(1)}$  and two-form $M_a^{i\;(2)}$  operators which satisfy the descent equations (\ref{covariantdescent}). It turns out that
\bea
&& M_a^{i\;(1)}=\mu^i_{a\; \jbar}\,d\,\phi^\jbar+  D_k\, \mu^i_{a\; \jbar}\, \rho^k\, \sigma^\jbar\nn\\
&& M_a^{i\; (2)}=D_k\, \mu^i_{a\; \jbar}\,\rho^k\,d\,\phi^\jbar +D_k\, \mu^i_{a\; \jbar}\,F^k\, \sigma^\jbar +\frac{1}{2}\,D_j\, D_k\, \mu^i_{a\; \jbar}\, \rho^j\,\rho^k\, \sigma^\jbar
\eea
Then
\bea
O_a = \bigl(M^{i\;(0)}_a +M^{i\;(1)}_a+M^{i\;(2)}_a\bigr)\,\bigl(\tilde{\phi}_i^* + \rho_i^* + F_i^*\bigr) 
\eea
is a $\delta$-cocycle and
\bea
&&O_a^{(0)}= \mu^i_{a\; \jbar}\,\sigma^\jbar\, F^*_i \nn\\
&&O_a^{(1)}= \mu^i_{a\; \jbar}\,\sigma^\jbar\, \rho^*_i +\bigl(\mu^i_{a\; \jbar}\,d\,\phi^\jbar+  D_k\, \mu^i_{a\; \jbar}\, \rho^k\, \sigma^\jbar\bigr)\,F^*_i \nn\\
&&O_a^{(2)}=\mu^i_{a\; \jbar}\,\sigma^\jbar\,\tilde{\phi}_i^*
+ \bigl(\mu^i_{a\; \jbar}\,d\,\phi^\jbar+  D_k\, \mu^i_{a\; \jbar}\, \rho^k\, \sigma^\jbar\bigr)\,\rho^*_i +\nn\\
&&\qquad +\bigl(D_k\, \mu^i_{a\; \jbar}\,\rho^k\,d\,\phi^\jbar +D_k\, \mu^i_{a\; \jbar}\,F^k\, \sigma^\jbar +\frac{1}{2}\,D_j\, D_k\, \mu^i_{a\; \jbar}\, \rho^j\,\rho^k\, \sigma^\jbar\bigr)\, F^*_i =\nn\\
&&\qquad  =\mu^i_{a\; \jbar}\,\sigma^\jbar\,\phi_i^*+ \bigl(\mu^i_{a\; \jbar}\,d\,\phi^\jbar+  \partial_k\, \mu^i_{a\; \jbar}\, \rho^k\, \sigma^\jbar\bigr)\,\rho^*_i +\nn\\
&&\qquad +\bigl(D_k\, \mu^i_{a\; \jbar}\,\rho^k\,d\,\phi^\jbar +\partial_k\, \mu^i_{a\; \jbar}\,F^k\, \sigma^\jbar +\frac{1}{2}\,D_j\, D_k\, \mu^i_{a\; \jbar}\, \rho^j\,\rho^k\, \sigma^\jbar +\nn\\
&&\qquad +\frac{1}{2}\,R^k _{ij;l}\,\rho^j\, \rho^l \,\mu^i_{a\; \jbar}\,\sigma^\jbar  \bigr)\, F^*_i 
\label{observablesfinal}
\eea
are observables of total fermionic number +2 satisfying the equivariant descent equations (\ref{equivariantdescent}).

This cocycle is $\gamma$-independent: therefore $O_a$ is also an observable of the matter theory.  This is possible since
the 2-form observable $O_a^{(2)}$ is, in fact, $G_\gamma$-invariant
\bea
G_\gamma \,O_a^{(2)}=0
\eea
Observables which satisfy this conditions are called ``chiral'', in the $N=2$ supersymmetric language.

\newsection{Varying the parameters of the BV action}

The observables described in the previous section are associated to  Beltrami differentials: one therefore 
expects that an  integrated 2-form $O^{(2)}_a$ is  related to the {\it holomorphic} derivative
of the BV action  with respect to the complex structure moduli $m^a$.  Since however the coordinates fields do depend
implicitly on the complex structure, this expectation is not completely realized: it turns out that the  integrated
observable  $O^{(2)}_a$ is not just given by the derivative of the BV action with respect to $m^a$ but it must
be supplemented with extra terms which make it invariant.  To explain this let us start from  the BV master equation
\bea
 && \sum_A\int S\phi^A\, S\, \phi_A^* (-1)^{A+1} =0
 \label{BVmaster}
 \eea
 where we denoted by $\phi^A$ and $\phi_A^*$ the fields and anti-fields of the model.  Let us now consider the variation of the
 master equation under  a  {\it generic} deformation $\delta$  of the parameters  --- which could be either the complex structure or metric --- on which $S$ depends:
  \bea
&& 0= \sum_A \int (\delta (S\phi^A))\, S\, \phi_A^* (-1)^{A+1} +S\phi^A\, (\delta (S\, \phi_A^* (-1)^{A+1})) =\nn\\
&&\qquad = \sum_A  \int (I \phi^A)\, S\, \phi_A^* (-1)^{A+1} +S\phi^A\, I\, \phi_A^* (-1)^{A+1}+\nn\\
&&\qquad\qquad  + \sum_A \int (S\,\delta \phi^A)\, S\, \phi_A^* (-1)^{A+1} +S\phi^A\, ( S\, \delta \phi_A^* (-1)^{A+1})=\nn\\
&&\qquad = \sum_A \int (I \phi^A)\, \frac{\partial \Gamma_{BV}}{\partial \phi^A} + \frac{\partial \Gamma_{BV}}{\partial \phi^*_A} \, I\, \phi_A^*  +  \sum_A \int (S\, \delta \phi^A)\, \frac{\partial \Gamma_{BV}}{\partial \phi^A} + \frac{\partial \Gamma_{BV}}{\partial \phi^*_A} \, (S\, \delta\, \phi_A^*) = \nn\\
&&\qquad =I\, \Gamma_{BV}+ \sum_A \int (S\, \delta \phi^A)\, \frac{\partial \Gamma_{BV}}{\partial \phi^A} + \frac{\partial \Gamma_{BV}}{\partial \phi^*_A} \, (S\, \delta\, \phi_A^*)=0
\label{BVmasterdeformed}
\eea
where we introduced, as in (\ref{Sdeformation}), the operator 
\bea I = \delta(S)
\eea 
of ghost number +1 which is the deformation of the BRST operator and which anti-commutes with $S$.  We also accounted for an implicit dependence of fields and anti-fields on the parameter
which are being varied. This is the case, of course, when we vary the complex structure in the B-model since the coordinates
fields depend on the complex structure in the way that has been computed in Section \ref{subsec:diffgeo}.

The second term in the last line of the equation (\ref{BVmasterdeformed}) above is 
$S$-trivial
\bea
&& \!\!\!\!  \!\!\!\!  \!\!\!\!  \!\!\!\!  \!\!\!\! \sum_A \int (S\,\delta \phi^A)\, S\, \phi_A^* (-1)^{A+1} +S\phi^A\, ( S\, \delta \phi_A^* (-1)^{A+1})=
S\, \int \bigl((S\,\delta \phi^A)\,  \phi_A^*  +S\phi^A\,  \delta \phi_A^*  \bigr)
\label{BVdeformedone}
\eea
Moreover
\bea
0 =\delta (S\, \Gamma_{BV}) =  I\, \Gamma_{BV} + S\, (\delta\,\Gamma_{BV})
\label{BVdeformedtwo}
\eea
Plugging both (\ref{BVdeformedone}) and (\ref{BVdeformedtwo}) into Eq. (\ref{BVmasterdeformed}) one obtains 
\bea
S\, \bigl(\delta\,\Gamma_{BV}- \int (S\,\delta \phi^A)\,  \phi_A^*  -S\phi^A\,  \delta \phi_A^*  \bigr)=0
\label{BVmasterdeformedthree}
\eea
This equation says that,  when (some of) the (anti-)fields depend implicitly on the deformation parameter, 
the BRST invariant observable associated to deformation parameter is not simply the variation of the action, but it 
must be supplemented with bilinear terms containing the anti-fields. The resulting BRST invariant integrated observable only depends
on the fermionic operator $I$ 
\bea
&& \hat{O}= \delta \Gamma_{BV} -\int \bigl((S\,\delta \phi^A)\,  \phi_A^*  +S\phi^A\,  \delta \phi_A^*\bigr)=\nn\\
&&\qquad = \sum_A \int  \delta(S\,\phi^A) \,\phi_A^*+S\,\phi^A \,\delta \phi_A^* -(S\,\delta \phi^A)\,  \phi_A^*  -S\phi^A\,  \delta \phi_A^*=\nn\\
&&\qquad =\sum_A \int  (I\,\phi^A) \,\phi_A^* 
\label{BVcovderivative}
\eea
The gauge-independent physics associated to the deformation $\delta$ is hence captured by the operator $I$. 
The operator $I$ satisfies the consistency condition
\bea
\{ I , S\} =0
\label{ISconsistency}
\eea
 Deformations $I$ which are $S$-commutators
 \bea
I = [S, L]
\label{BVtrivialI}
\eea
correspond to integrated observables $\hat{O}$ which are trivial in the BV sense
\bea
&& \hat{O}=  \int \sum_A [S, L]\,\phi^A \,\phi_A^*=S\,  \sum_A \int L\phi^A\, \phi_A^*
\eea 
In the BV framework,  gauge-fixing is performed by choosing a gauge-fermion  functional $\Psi[\phi^A]$ 
and putting
\bea
\phi^*_A = \frac{\delta}{\delta \phi^A}\, \int \Psi[\phi^A]
\label{BVgf}
\eea
From (\ref{BVcovderivative}) it follows that to the BV observable $\hat{O}$ there corresponds the  gauge-fixed integrated observable 
\bea
\hat{O}^{g.f} =   I\, \int  \Psi(\phi^A)
\eea 
$\hat{O}^{g.f}$ is BRST closed modulo the equations of motion generated by $I$
\bea
S\, \hat{O}^{g.f} = - I \, \Gamma^{g.f.}
\eea
When $I$ is a $S$ commutator, as in (\ref{BVtrivialI}),   the corresponding gauge-fixed observable is BRST-trivial modulo the equations of motions associated to $L$:
\bea
\hat{O}^{g.f}=  S\, \int  (L \Psi[\phi^A]) - L\, \Gamma^{g.f.}[\phi^A]
\label{GFtrivialL}
\eea
We see therefore that deformations which generate $I$ which are $S$-commutators, do {\it not} necessarily decouple in physical
correlators:  Eq. (\ref{GFtrivialL}) says that the insertion of a trivial $I$ operator in a physical correlator gives contact terms generated by the operator $L$. Those contact   may or may not vanish according to the specific form of both $L$ and the physical observables. In the following we will determine the operator $L$ for different BRST trivial deformations of the B-model, to assess, in a gauge-independent  way, their decoupling --- or lack thereof. 

\subsection{Dependence on the target space metric}

Let us denote by $\delta_{\delta g}$ the variation of the target space metric $g_{i\jbar} \to g_{i\jbar} +\delta g_{i\jbar}$  In this case the (anti-)fields are left
invariant by the deformation and therefore the corresponding operator insertion is just given 
by the variation of the BV action\footnote{We included in this formula for completeness also the trivial $(\theta_i, H_i)$  sector.} :
\bea
&& \hat{O}_{\delta g} \equiv \delta_{\delta g}\Gamma_{BV} = \int  \bigl[\delta_{\delta g} ( \hat{S}\,F^i)\, F_i^*+ \delta_{\delta g} ( \hat{S}\,H_i)\, H^{*\,i}+\nn\\
&&\qquad - \delta_{\delta g}\,\Gamma^i _{ i_\gamma(\rho);j}\,\rho^j \, \rho_j^*-  \delta_{\delta g}\,\Gamma^i _{ i_\gamma(\rho);j}\,F^j \, F_j^* + \delta_{\delta g}\,\Gamma^i _{ i_\gamma(\rho);j}\,\theta_i \, \theta^{j\;*} + \delta_{\delta g}\,\Gamma^i _{ i_\gamma(\rho);j}\,H_i \, H^{j\;*}\bigr]=\nn\\
&&\qquad = \int \Bigl[\bigl( \delta\Gamma^i_{d\phi\,j}\,\rho^j+\frac{1}{2}\, \partialbar_\sigma \,\delta_{\delta g}\Gamma^i _{\rho;j}\,  \rho^j  +\frac{1}{3}\, \bigl(D_{i_\gamma(\rho)}\, \delta_{\delta g}\,\Gamma^i _{ \rho;\rho} - D_{\rho}\, \delta_{\delta g}\,\Gamma^i _{ i_\gamma(\rho);\rho}  \bigr)\, F_i^* +\nn\\
&&\qquad +\bigl( \delta_{\delta g}\Gamma^j_{i_\gamma(d\phi)\; i}\,\theta_j -\bigl(\partialbar_\sigma \,\delta_{\delta g}\,\Gamma^j_{i_\gamma(\rho)\; i}+D_{i_\gamma(\rho)}\,\delta_{\delta g}\,\Gamma^j_{i_\gamma(\rho)\; i}\bigr)\,\theta_j\bigr)\, H^{*\,i}+\nn\\
&&\qquad - \delta_{\delta g}\,\Gamma^i _{ i_\gamma(\rho);j}\,\rho^j \, \rho_j^*-  \delta_{\delta g}\,\Gamma^i _{ i_\gamma(\rho);j}\,F^j \, F_j^* + \delta_{\delta g}\,\Gamma^i _{ i_\gamma(\rho);j}\,\theta_i \, \theta^{j\;*} + \delta_{\delta g}\,\Gamma^i _{ i_\gamma(\rho);j}\,H_i \, H^{j\;*}\Bigr]=\nn\\
&&\qquad =S\, \int ( \frac{1}{2}\,\delta_{\delta g} \Gamma_{\rho \rho}^i \, F_i^* -\delta_{\delta g}\,\Gamma^j_{i_\gamma(\rho)\; i}\, \theta_j\,H^{*\,i})
\label{kahlerBV}
\eea 
This shows that a deformation of the target space metric is BRST trivial in the space of fields and anti-fields and  that 
$I_{\delta g} = \delta_{\delta g} (S)$  is a BRST commutator
\bea
I_{\delta g}= [S, L_{\delta g}]
\eea
where the operator $L_{\delta g}$ acts non-trivially only on the fields $F^i$ and $H_i$
\bea
&& L_{\delta g} \ F^i=  \frac{1}{2}\,\delta_{\delta g} \Gamma_{j k}^i\, \rho^j\, \rho^k\nn\\
&& L_{\delta g}\, H_i =  -\delta_{\delta g}\,\Gamma^j_{k i}\, i_\gamma(\rho^k)\,\theta_j
\label{Lmetric}
\eea
Since  $\hat{O}_{\delta g}$  is trivial in  the space of  both fields and anti-fields,  it follows that, upon gauge fixing, the corresponding insertion is BRST trivial {\it up to terms proportional to the equations of motions generated by $L_{\delta g}$}:  From (\ref{Lmetric}) these are the equations of motion of the auxiliary  fields $F^i$ and $H_i$.   If one therefore considers correlators involving only  0-form observables (\ref{observablesfinal}) associated to the complex structure moduli,  
\bea
O^{(0)}_a = \mu^i_{a\; \jbar}\,\sigma^\jbar\, F^*_i 
\label{0formpicture}
\eea
these contact terms vanish, since the observables do not depend on either $F^i$ or $H_i$.  Correlators of such observables are
therefore independent of the target space metric.

\subsection{Dependence on the holomorphic complex structure moduli}

Since the coordinate
fields $\phi_i$ and anti-fields $\phi_\ibar^*$  depend implicitly on the complex structure moduli, the holomorphic
derivative  $\partial_a\,\Gamma_{BV}$ of the BV action differs from the integrated BRST-invariant observable  $\hat{O}_a$, as specified in (\ref{BVcovderivative})
\bea
&& \int \hat{O}_a^{(2)}= \partial_a \, \Gamma_{BV} -  \int [(S\,\delta \phi^A)\,  \phi_A^*  +S\phi^A\,  \delta \phi_A^*]=\nn\\
&&\qquad  \qquad =\int (I_a\phi^i\,\phi_i^* +I_a\rho^i\,\rho_i^*+ I_a\,F^i\,F_i^* + I_a\phi^\ibar\, \phi_\ibar^*+  I_a\sigma^\ibar\, \sigma_\ibar^*)
\eea
where we neglected the BRST trivial  doublet $(\theta_i, H_i)$ which gives an equally trivial contribution to the observable.

To compute the corresponding $I_a = \partial_a (S)$ we need to specify the implicit dependence on the holomorphic moduli of
fields  and anti-fields. We discussed the dependence on the complex structure moduli of the coordinate fields $(\phi^i, \phi^\ibar)$
in Section \ref{subsec:diffgeo}: it is given by the Beltrami parametrization (\ref{beltramione}), which also determines the dependence
on the complex structure moduli of connections, curvature tensors  and anti-fields as shown in  Eqs. (\ref{holoconnection}), (\ref{holocurvatures}) and (\ref{holoantifields}).

The fields other than the coordinates  take values on non-compact affine field spaces with no boundaries. Since they are integrated over in the functional integral that defines quantum averages, their specific dependence on the moduli is, in fact, irrelevant  for the computation of normalized quantum correlators.  It is therefore convenient to choose a dependence for the (anti-)fields other than the coordinate fields, which is explicitly covariant under target space
holomorphic reparametrization, i.e.
\bea
D_a \rho^i = D_a \, F^i = D_a \theta_i = D_a\, H_i=0
\eea
and analogously for the corresponding anti-fields. With this choice
\bea
&& I_a \phi^i =- \mu^i_{a\;\jbar} \,\sigma^\jbar\nn\\
&& I_a \rho^i =\mu^i_{a\;\ibar}\,d\phi^\ibar  - \partial_j\,\mu_{a\;\sigmabar}^i \,\rho^j + I_{D_a g}\, \rho^i \nn\\
&&I_a \, F^i = d\phi^\kbar\, D_\rho \,\mu_{a\;\kbar}^i  - \frac{1}{2}\,D_\rho^2\,\mu^{i}_{a \bar{\sigma}}+\frac{1}{2}\,R^i_{\rho l;\rho}\,\mu^{l}_{a \bar{\sigma}}- \partial_j \mu^i_{a\; \sigma}\, F^j +I_{D_a g}\, F^i \nn\\
&& I_a\, \phi^\ibar=0\nn\\
&& I_a\, \sigma^\ibar=0
\label{Iafinal}
\eea
where $I_{D_a g}$ is the deformation of the BRST under a change of the target space metric
\bea
g_{i\jbar} \to g_{i\jbar} + D_a\, g_{i\jbar}
\eea
We have just shown that  $I_{D_a g}$ is an $S$-commutator and that the associated insertion is BRST trivial.  Therefore, by comparing
(\ref{Iafinal}) with  (\ref{observablesfinal}),  we conclude that the integrated  observable associated to the deformation $I_a$ is, up to BRST trivial terms, precisely the one  which descends from  the 0-form  operator (\ref{0formpicture}).

The insertion of an  integrated observable $\hat{O}_a$  in a BRST invariant correlator is related to the holomorphic derivative $\partial_a$ of the same correlator, but does not coincide with it.  The basic reason for this is that the observable is  BRST closed  only up to
terms which are proportional to the equations of motion.  The consequence of this is that the insertion of  $\hat{O}_a$ is obtained  by taking an appropriate {\it covariant} derivative of the correlator,   whose connection is fixed by BRST invariance. To see this let us
consider the holomorphic derivative $\partial_a$ of a correlator involving  another  integrated observable $\hat{O}_b$
\bea
&&\partial_a \, \langle \hat{O}_b\rangle = \langle \bigl[ (\partial_a\, S\, \int \Psi)\, (\int I_b \,\Psi) + \partial_a I_b\int\Psi\bigr]\rangle=\nn\\
&&\qquad = \langle \bigl[ (S\, \int \partial_a\Psi + \int I_a \Psi)\,( \int I_b \,\Psi) + \partial_a(I_b)\int\Psi+I_b\int \partial_a \Psi\bigr]\rangle=\nn\\
&&\qquad =\langle  \int I_a \Psi\,  \int I_b \,\Psi - \int   \partial_a\Psi\, I_b\, \Gamma  + \partial_a(I_b)\int\Psi+I_b\int \partial_a \Psi\bigr]\rangle=\nn\\
&&\qquad =\langle  \int I_a \Psi\,  \int I_b \,\Psi  +I_{ab}\int\Psi\rangle
\eea
where in the last line  we introduced the fermionic operator $I_{ab}$ symmetric in $a$ and $b$:
\bea
I_{ab} = \partial_a\, I_b = \partial_a\,\partial_b\, (S) = - D_a\, \mu_{b\; \jbar}^i\,\sigma^\jbar \, \partial_i + \cdots
\eea
Here the  the dots denote the action of the operator on fields other than the coordinate fields $\phi^i$ and $\phi^\ibar$. 
 We see that the correlator of two BRST invariant integrate observable writes as 
\bea
\langle \hat{O}_a \hat{O}_b\rangle = \partial_a \, \langle \hat{O}_b\rangle - \langle \int I_{ab}\, \Psi\rangle
\eea
The last term is a local integrated operator which encodes the contact between the two local operators $ \hat{O}_a$ and $\hat{O}_b$.  We can think of the contribution of  $I_{ab} \, \Psi$ as a renormalization counterterm which must be added to the correlator of two integrate observables to make it BRST-invariant --- i.e. gauge-independent. The overall effect of this contact term is   that the insertion of $\hat{O}_a$ is given by  taking a  covariant derivative
\bea
\langle \hat{O}_a \hat{O}_b\rangle = \mathcal{D}_a({\Gamma})\,  \langle \hat{O}_{b}\rangle
\eea
build with a connection $\Gamma_{ab}^c$ in the moduli space which is determined from $I_{ab}$. 
This connection, in the K\"ahler case, is precisely the connection compatible with the Zamolodchikov metric on the moduli
space of complex structure.  For a detailed derivation of these statements  see S. Giusto, Ph. D. dissertation thesis \cite{Giusto:1999}.
The computation of this connection in the non-K\"ahler case is left to future work.

\subsection{The dependence on the anti-holomorphic complex structure moduli}

Let us now turn to consider the  BRST-invariant operator insertion associated to the anti-holomorphic derivative  of the BV action with respect to the complex structure:
The anti-holomorphic derivative of the simple BV action, neglecting again for simplicity the $(\theta_i, H_i)$ sector, is
\bea
&& \int \hat{O}_\abar^{(2)}= \int (I_\abar\phi^i\,\phi_i^* +I_\abar\rho^i\,\rho_i^*+ I_\abar\,F^i\,F_i^* + I_\abar\phi^\ibar\, \phi_\ibar^*+  I_\abar\sigma^\ibar\, \sigma_\ibar^*)
\eea
where the deformation $I_\abar= \partial_\abar(S)$ is
\bea
&& I_{\bar a} \phi^i = 0 \ , \nn \\
&& I_{\bar a} \rho^i =I_{D_\abar g}\, \rho^i - \frac{1}{2}\, D^i (\mubar_{\abar\; jk})\, i_\gamma(\rho^j\, \rho^k)\nn \\
&& I_{\bar a} F^i = I_{D_\abar g}\, F^i-D^i\,\mubar_{\abar\; jk}\, i_\gamma(\rho^j)\, F^k-
D^i\,\mubar_{\abar\; k j}\,d\phi^k\, \rho^j +\nn\\
&&\qquad + \frac{1}{2}\,  D_\ibar \,D^i (\mubar_{\abar\; jk}) \sigma^\ibar\, \rho^j\,\rho^k + \frac{1}{3}\,\mathcal{R}_{\abar\;jk;l}^i\,  i_\gamma(\rho^j)\, \rho^k\,\rho^l  \nn \\
&& I_{\bar a} \phi^\ibar = -\mubar^\ibar_{\abar\;i}\, i_\gamma(\rho^i) \ , \nn \\
&& I_{\bar a} \sigma^\ibar = \mubar^\ibar_{\abar\; i}\, i_\gamma(d\phi^i) + \partialbar_{\kbar} \mubar_{\abar\; i}^\ibar\, \sigma^\kbar\,i_\gamma(\rho^i) 
\eea
where $\mathcal{R}_{\abar\;jk;l}^i$ is the tensor defined in  (\ref{Rmutensor}), which vanishes for K\"ahler metrics, while $I_{D_\abar g}$ is the deformation of $S$ associated to a shift of the metric
\bea
g_{i\jbar}\to g_{i\jbar} + D_\abar g_{i\jbar}
\eea
The anti-holomorphic deformation $I_\abar= \partial_\abar(S)$ turns out to be a $S$ commutator
\bea
I_\abar = [S, L_\abar +L_{D_\abar g}]
\eea
where $L_\abar$  is a bosonic operator  whose non-trivial action is
\bea
&&  L_\abar \, \sigma^\ibar = \mubar^\ibar_{\abar\; i}\, i_\gamma(\rho^i)\nn\\
&& L_\abar \, F^i =\frac{1}{2}\,D^i\,\mubar_{\abar\; jk}\, \rho^j\, \rho^k
\label{Labarfinal}
\eea
Therefore the integrated anti-holomorphic insertion is BRST  trivial in the BV sense
\bea
&& \int \hat{O}_\abar^{(2)}=S\, \int (L_\abar \, \sigma^\ibar\, \sigma_\ibar^* + L_\abar \, F^i \, F_i^*)  +\nn\\
 &&\qquad +S\, \int ( \frac{1}{2}\,\delta_{D_\abar g} \Gamma_{\rho \rho}^i \, F_i^* -\delta_{D_\abar g}\,\Gamma^j_{i_\gamma(\rho)\; i}\, \theta_j\,H^{*\,i})
\eea
The terms in the second line of the r.h.s is a trivial term associated to a target space deformation of the metric, which, as we discussed above, does decouple when inserted in a correlator of holomorphic deformations.  
The terms in the first line, instead, correspond, upon gauge-fixing, to an insertion which is  BRST trivial up to terms proportional to the equations of motions generated by $L_\abar$:  from (\ref{Labarfinal}), this means   trivial up to terms proportional to  equations of motion of 
$\sigma^\ibar$. Since  0-form observables (\ref{observablesfinal}) associated to the complex structure moduli,  
\bea
O^{(0)}_a = \mu^i_{a\; \jbar}\,\sigma^\jbar\, F^*_i 
\label{0formpicturebis}
\eea
do depend on $\sigma^\ibar$, the anti-holomorphic integrated insertion does not decouple when inserted in a correlator containing holomorphic observables: the terms proportional to the equation of motions of $\sigma^\ibar$ produce non-vanishing contact terms
when observables like (\ref{0formpicturebis}) are present in a quantum correlator.  This is the root of the anti-holomorphic dependence of integrated correlators of observables $O^{(0)}_a$, i.e. of topological strings amplitudes. Let us note that the contacts produced by anti-holomorphic insertions are proportional to the antighost fields $\gamma^\alpha$: in the ``matter''  formulation in which $\gamma^\alpha$ is ignored, the contact would be ---  incorrectly --- interpreted as a BRST anomaly.  

\section*{Acknowledgments}

It is a great pleasure to thank C. Becchi and S. Giusto for earlier collaboration with me on this topic:  most of what I have presented
in this article I understood as a result of working with them. I am also grateful to A. Tomasiello for sharing with me his work with A. Kapustin on non-K\"ahlerian B-models. As noted, trying to solve their puzzle provided the original inspiration this paper. I am also greatly indebted to D. Rosa for patiently and frequently discussing the details of many puzzles and confusing moments with me over the course of the last few months. 
 
Finally, I would like express my gratitude to Raymond Stora, who, during the time I had the fortune to know him, generously  and graciously bestowed upon me not only his profound knowledge of both physics and mathematics, but also his enthusiasm for everything beautiful, his inexhaustible intellectual curiosity and his exquisite  humanity. 

The work of CI was supported in part by  INFN, by Genoa University Research Projects (P.R.A.) 2014 and 2015. 

\newpage

\providecommand{\href}[2]{#2}


\begin{thebibliography}{10}


\bibitem{Labastida:1988zb} 
  J.~M.~F.~Labastida, M.~Pernici and E.~Witten,
  ``Topological Gravity in Two-Dimensions,''
  Nucl.\ Phys.\ B {\bf 310}, 611 (1988).

  E.~Witten, ``On the Structure of the Topological Phase of Two-dimensional Gravity,''
  Nucl.\ Phys.\ B {\bf 340}, 281 (1990).

\bibitem{Ouvry:1988mm}
  S.~Ouvry, R.~Stora and P.~van Baal,
  ``On the Algebraic Characterization of Witten's Topological Yang-Mills
  Theory,''
  Phys.\ Lett.\  B {\bf 220} (1989) 159. 

\bibitem{Baulieu:1988xs}
  L.~Baulieu and I.~M.~Singer, ``Topological Yang-Mills Symmetry,''  Nucl.\ Phys.\ Proc.\ Suppl.\  {\bf 5B} (1988) 12.

  L.~Baulieu and I.~M.~Singer,
  ``The Topological Sigma Model,''
  Commun.\ Math.\ Phys.\  {\bf 125} (1989) 227.

  H.~Kanno,
  ``Weyl AlgebraStructure and Geometrical Meaning of the BRST transformation in Topological Quantum Field
  Theory,"
  Z.\ Phys.\  C {\bf 43} (1989) 477.

\bibitem{Montano:1988dr} 
  D.~Montano and J.~Sonnenschein,
  ``Topological Strings,''
  Nucl.\ Phys.\ B {\bf 313}, 258 (1989).

\bibitem{Becchi:1995ik} 
  C.~M.~Becchi and C.~Imbimbo,
  ``Gribov horizon, contact terms and Cech-De Rham cohomology in 2-D topological gravity,''
  Nucl.\ Phys.\ B {\bf 462}, 571 (1996)
[\href{http://arxiv.org/abs/hep-th/9510003}{{\tt arXiv:hep-th/9510003}}].

\bibitem{Witten:1988xj} 
  E.~Witten,
  ``Topological Sigma Models,''
  Commun.\ Math.\ Phys.\  {\bf 118}, 411 (1988).
  E.~Witten,
  {\sl Mirror manifolds and topological field theory},
  In *Yau, S.T. (ed.): Mirror symmetry I* 121-160
 [\href{http://xxx.lanl.gov/abs/hep-th/9112056}{{\tt arXiv:hep-th/9112056}}].


\bibitem{Bershadsky:1993ta} 
  M.~Bershadsky, S.~Cecotti, H.~Ooguri and C.~Vafa,
  Nucl.\ Phys.\ B {\bf 405}, 279 (1993)
   [\href{http://xxx.lanl.gov/abs/hep-th/9302103}{{\tt arXiv:hep-th/9302103}}].

  M.~Bershadsky, S.~Cecotti, H.~Ooguri and C.~Vafa,
  Commun.\ Math.\ Phys.\  {\bf 165}, 311 (1994)
   [\href{http://xxx.lanl.gov/abs/hep-th/9309140}{{\tt arXiv:hep-th/9309140}}].

\bibitem{Bae:2015eoa} 
  J.~Bae, C.~Imbimbo, S.~J.~Rey and D.~Rosa,
  ``New Supersymmetric Localizations from Topological Gravity,''
  JHEP {\bf 1603}, 169 (2016)
  [\href{http://arxiv.org/abs/1510.00006}{{\tt arXiv:1510.00006}}].
  
  C.~Imbimbo and D.~Rosa,
  ``Topological anomalies for Seifert 3-manifolds,''
  JHEP {\bf 1507}, 068 (2015)
  [\href{http://arxiv.org/abs/1411.6635}{{\tt arXiv:1411.6635}}].

\bibitem{Giusto:1999}
S.~Giusto, ``Topological Field Theories and Strings'', Ph. D Thesis, University of Genoa, 1999.

\end{thebibliography}
\end{document}